\magnification=\magstep1        
\looseness=2
\hsize=6.7 truein
\vsize=9.0 truein
\baselineskip=20pt plus2pt
\parskip=6pt plus4pt minus2pt

\font\medfont=cmssbx10 scaled \magstep2

% -----------------------------auto numbering of equations
%                    \nxt{\name_for_equation} - sets the name and prints (N)
%                    \Eqn{\name_for_equation} - cites the equation
\newcount\qno  \qno=0
\def\Eqnew#1{\global\advance\qno by 1%
             \xdef#1{\number\qno}}
\def\Eqn#1{\leqno(#1)}
\def\nxt#1{\Eqnew{#1}\Eqn{#1}}
% --------------------------------------auto numbering of plots
%                    \NewPlot{\name_for_plot} - sets the name and prints N
%                    \Plot{\name_for_plot}    - cites the plot
\newcount\plotnumber  \plotnumber=0%
\def\Pnew#1{\global\advance\plotnumber by 1%
             \xdef#1{\number\plotnumber}}%
%
%

%----------------------------------------------------------
%                             macros to define the reference formats
%                    \NewRef{\name_for_ref} - sets the name and prints [N]
%                    \Plot{\name_for_plot}    - cites the plot
\newcount\refnumber  \refnumber=0%
\def\Rnew#1{\global\advance\refnumber by 1%
             \xdef#1{\number\refnumber}}%
\def\Ref#1{[#1\thinspace]}%
\def\ref#1{#1\thinspace}%
\def\Reff#1{(#1\thinspace)}%
\let\nr=\NewRef
\def\nrr#1{\Rnew{#1}\ref{#1}}%
  \def\ps{\noindent\goodbreak
      \parshape 2 0truecm 6.7truein 0.5truein 6.5truein}
  \def\\{\hfil\break}
  \def\aa{{\it Astron. Astrophys.}\rm}
  \def\aj{{\it Astron. J.~}\rm}
  \def\apj{{\it Astrophys. J.~}\rm}
  
  \def\apjs{{\it Astrophys. J. Suppl.~}\rm}
  \def\mnras{{\it Mon. Not. R.astr. Soc.~}\rm}
%-------------------------------------------

\def\Msun{~M_{\odot}\ }

\def\Mpch{h^{-1}{\rm Mpc}}

\def\etal{{\it et al.}\ }

\def\M10{{\times 10^{10} M_{\odot}\ }}

%::::::::::::::::::::::::::::::::::::::::::::::::::::::::::::::::::::::::::::::
%       Define a decent looking underline.
%::::::::::::::::::::::::::::::::::::::::::::::::::::::::::::::::::::::::::::::

\def\underule#1{$\setbox0=\hbox{#1} \dp0=\dp\strutbox
    \m@th \underline{\box0}$}
%::::::::::::::::::::::::::::::::::::::::::::::::::::::::::::::::::::::::::::::
%       Define sequences for "less than or approximately equal to" and
%       "greater than or approximately equal to"
%::::::::::::::::::::::::::::::::::::::::::::::::::::::::::::::::::::::::::::::

\def\applt{\mathrel{\mathpalette\@versim<}}
\def\appgt{\mathrel{\mathpalette\@versim>}}
\def\@versim#1#2{\lower2pt\vbox{\baselineskip0pt \lineskip-.5pt
   \ialign{$\m@th#1\hfil##\hfil$\crcr#2\crcr\sim\crcr}}}

%::::::::::::::::::::::::::::::::::::::::::::::::::::::::::::::::::::::::::::::
%       Define a sequence for obtaining a vector symbol with a hook.
%::::::::::::::::::::::::::::::::::::::::::::::::::::::::::::::::::::::::::::::

{\catcode`p=12 \catcode`t=12 \gdef\\#1pt{#1}}
\let\getfactor=\\
\def\kslant#1{\kern\expandafter\getfactor\the\fontdimen1#1\ht0}
\def\vector#1{\ifmmode\setbox0=\hbox{$#1$}%
    \setbox1=\hbox{\the\scriptscriptfont1\char'52}%
	\dimen@=-\wd1\advance\dimen@ by\wd0\divide\dimen@ by2%
    \rlap{\kslant{\the\textfont1}\kern\dimen@\raise\ht0\box1}#1\fi}

\null
\centerline{\medfont Numerical Simulations in Cosmology I}
\bigskip
\centerline{\bf Anatoly Klypin}

\centerline{\it New Mexico State University, Department of Astronomy,
Las Cruces, NM 88001, USA}
\bigskip
\centerline{\bf Abstract}
The purpose of these lectures is to give a short introduction into a
very vast field of numerical simulations for cosmological
applications. I focus on major features of the simulations: the
equations, main numerical techniques, effects of resolution, and
methods of halo identification.

\bigskip
\centerline{\bf 1. Introduction}

Numerical simulations play a very significant role in cosmology. It
all started in 70s with simple N-body problems solved using primitive
N-body codes with few hundred particles. Later the code which we now
call Particle-Particle code (direct summation of all two-body forces)
was polished and brought to the state-of-art \nr{\Aarseth}. Already
those early efforts brought some very valuable fruits. In 1970 Peebles
\nr{\Peeblesa} studied collapse of a cloud of particles as a model of cluster
formation. The model had 300 points initially distributed within a
sphere with no initial velocities. After the collapse and
virialization the system looked like a cluster of galaxies. Those
early simulations of cluster formation, though producing cluster-like
objects, signaled the first problem -- simple model of initially
isolated cloud (top-hat model) results in the density profile of the
cluster which is way too steep (slope -3 -4) as compared with real
clusters. The problem was addressed by Gunn \& Gott \nr{\GunnGott},
who introduced a notion of secondary infall in an effort to solve the
problem.  Another keystone work of those times is the paper by White
\nr{\Whitea}, who studied collapse of 700 particles with different
masses. It was shown that if one distributes the mass of a cluster to
individual galaxies (=points), two-body relaxation will result in mass
segregation which is not compatible with observed cluster. This was
another manifestation of dark matter in clusters. This time it was
shown that the dark matter of a cluster is not in galaxies.  Survival
of substructures in galaxy clusters was another problem addressed in
the paper. It is quite remarkable that 20 years later we are still
facing the same two problems. For numerical models (at least some of
them) it is the excessive mass segregation, and, thus, a wrong
structure of objects due to insufficient number of particles. For
cluster dynamics it is the amount of substructure (buzzwords:
``merging'', ``hydrostatic equilibrium'').

First cosmological simulations (growth and collapse of initially small
fluctuations in expanding Universe) were done at about the same time
(mid 70s). An example: Aarseth, Gott \& Turner \nr{\AarsethGT} studied
the evolution of fluctuations using 5000 particles moving in expanding
sphere. The particles were initially placed at random within the
sphere (thus, the flat $n=0$ spectrum of initial fluctuations). The
sphere was needed to handle boundary conditions: if a particle happens
to hit the sphere it is reflected back like a ball in a billiard. At
that time the correlation function completely dominated cosmological
landscape. Simple hierarchical clustering arguments \nr{\Peeblesb}
indicated that in order to reproduce the observed slope of the
correlation function of galaxies $\gamma =-1.77$, the spectrum of
initial fluctuations should be flat. Results of N-body simulations
seems to confirm the conclusion. Another result, which followed from
the same arguments was used to defeat a cosmological scenario proposed
by Zeldovich \nr{\Zeldovich}, which was called at that time
``adiabatical fluctuations'' and later resurfaced under the name of
the Hot Dark Matter.  Because the Zeldovich model suggested very
distinct scale in the initial spectrum of fluctuations (almost no
fluctuations on small scales), the argument was that it should not
produce a scale-free power-law for the correlation function. As we
know now, both arguments were wrong. And it was found thanks to
numerical simulations.  Efstathiou \& Eastwood \nr{\EfEastwood} using
much better code (one of the first results with
Particle-Particle-Particle-Mesh (P$^3$M) code) with 20,000 particles
showed that the model with the flat spectrum (initial random
distribution) fails to produce power-law correlation function. Klypin
\& Shandarin \nr{\KlypinShandarin} showed that the Hot-Dark-Matter
(HDM) model predicts the power-law for the correlation function. The
HDM model is dead anyway (it does not have enough high-$z$ objects),
but the history of the model is really amazing: it was killed for sure
so many times ... because of wrong reasons.

Generation of initial condition with given amplitude and spectrum of
fluctuations was a problem for some time. The only correctly simulated
spectrum was the flat spectrum which was generated by randomly
distributing particles. In order to generate fluctuations with power
spectrum, say $P(k) \propto k^{-1}$, Aarseth, Gott \& Turner
\Ref{\AarsethGT} placed particles along rods. Formally, it generates
the spectrum, but the distribution has nothing to do with cosmological
fluctuations.  As far as I know, the papers of Doroshkevich \etal
\nr{\Doroshkevich} for two dimensions and Klypin \& Shandarin
\Ref{\KlypinShandarin} in three dimensions were the first where
initial conditions were generated using the Zeldovich \Ref{\Zeldovich}
approximation -- the method which since then was used to generate
initial conditions.

Starting mid 80s the field of numerical simulations is blooming: new
numerical techni\-ques were invented, old ones were perfected, many
publications (and, occasionally, results) are based on numerical
modeling.  To large extend, this have changed our way of doing
cosmology.  Instead of questionable assumptions and waving-hands
arguments, we have tools of testing our hypothesis and models. As an
example, I mention two analytical approximations which were validated
by numerical simulations. The importance of both approximations is
difficult to overestimate. The first is the Zeldovich approximation,
which paved the way of understanding the large-scale structure of the
galaxy distribution. The second is the Press-Schechter \nr{\PS}
approximation, which gives the number of objects formed at different
scales at different epochs. Both approximations cannot be formally
proved. The Zeldovich approximation formally is not applicable for
hierarchical clustering. It must start with smooth perturbations
(truncated spectrum). Nevertheless, numerical simulations have shown
that even for the hierarchical clustering the approximation can be
used with appropriate filtering of initial spectrum (e.g.
[\nrr{\ColesMS},\nrr{\Sath},\nrr{\BCM}]). The Press-Schechter
approximation is also difficult to justify without numerical
simulations. It operates with initial spectrum and linear theory, but
then (very long jump in assumptions) it predicts the number of objects
at very nonlinear stage. Because it is not based on any realistic
theory of nonlinear evolution (we just do not have one), it was an
ingenious, but a wild guess. If anything, the approximation is based
on a simple spherical top-hat model. But simulations show that objects
do not form in this way -- they are formed in a complicated fashion
through multiple mergers and accretion along filaments. Still this
very simple (and a very useful) prescription gives quite accurate
predictions
\nr{\EfRees}.  

The following of this lecture is organized in the following way.
Section 2 gives the equations which we solve to follow the evolution
of initially small fluctuations. A brief discussion  of different
methods is given in section 3. Effects of the resolution and some
other technical details are also discussed in Section 3. Identification of
halos (``galaxies'') is discussed in Section 4.
\bigskip
\centerline{\bf 2. Equations of evolution of fluctuations in an expanding
universe} 

Usually the problem of the formation and dynamics of cosmological
objects is formulated as N-body problem: for N point-like objects with
given initial positions and velocities find their positions and
velocities at any consequent moment. It should be remembered that this
just a short-cut in our formulation -- just to make things simple and
avoid any discussion. While it still mathematically correct in many
cases, it does not explain what we are doing. If we are literally to
take this approach, we should follow the motion of zillions of axions,
baryons, neutrinos, and whatever else our Universe is made of. So,
what it has to do with the motion of those few millions of particles
in our simulations? The correct approach is to start with the Vlasov
equation coupled with the Poisson equation (and proper initial and
boundary conditions). If neglect the baryonic component, which of
course is very interesting, but would complicate our situation too
much, the system is described by distribution functions $f_i({\bf x},
{\bf\dot x}, t)$ which should include all different clustered
components $i$. For a simple CDM model we have only one component
(axions or whatever it is). For more complicated Cold plus Hot Dark
Matter (CHDM) with few different types of neutrinos the system
includes one DF for the cold component and one DF for each type of
neutrino (\nr{\KHPR}). The equations for the evolution of $f_i$ are:
$$\eqalign{
&{\partial f_i \over \partial t} + {\bf x}{\partial f_i\over \partial{\bf x}}
	-  \nabla \phi{\partial f_i\over \partial{\bf\dot x}} =0, \quad
\nabla^2\phi = 4\pi G a^2 (\rho({\bf x}, t) -\rho_b(t))
              = 4\pi G a^2\Omega_{\rm dm}\delta_{\rm dm}, \cr
&\delta_{\rm dm}({\bf x}, t) = (\rho_{\rm dm}-\langle\rho_{\rm
	 dm}\rangle)/\langle\rho_{\rm dm}\rangle),\quad
\rho_{\rm dm}({\bf x}, t) =a^{-3}\sum_i m_i\int d^3{\dot x}
f_i({\bf x},{\bf\dot x},t). \cr }
\nxt{\EqVlasov}
$$
Here $a$ is the expansion parameter, $\Omega_{\rm dm}$ is the
contribution of the clustered dark matter to the mean density of the
Universe, $m_i$ is the mass of a particle of $i$th component of the
dark matter. The solution of the Vlasov equation can be given in terms
of characteristic equations, which {\it look} like equations of
particle motion. The distribution function is constant along each
characteristic. Complete set of characteristics is equivalent to the
Vlasov equation. Of course, we can not have the complete set, but we
can follow the evolution of the system (with some accuracy) if we
select a representative sample of characteristics. One way of doing
this would be to split initial phase space into small domains, take
only one characteristic as representative for this volume element, and
follow the evolution of the system of the ``particles'' in
self-consistent way. In models with one ``cold'' component of
clustering dark matter (like CDM or $\Lambda$CDM) the initial velocity
is a unique function of coordinates (only ``Zeldovich'' part is
present, no thermal velocities). This means that we need to split only
coordinate space, not velocity space. For complicated models with
significant thermal component (CHDM), the distribution in full phase
space should be taken into account. Usually, this is simulated by
placing 1-10 particles in each coordinate with velocities, which mimic
the distribution of real ``thermal'' velocities. Depending on what we
are interested in, we might split initial space into equal-size boxes
(typical setup for PM or P$^3$M simulations) or we could divide some
area of interest (say, where a cluster will form) into smaller boxes,
and use much bigger boxes outside the area (to mimic gravitational
forces of the outside material). In any case, the mass assigned to a
``particle'' is equal to the mass of the domain it represents. Now we
can think of the ``particle'' either as a small box, which moves with
the flow, but does not change its original shape, or as a point-like
particle. Both presentations are used in simulations. None is superior
to another.

There are different forms of final equations. If we choose
``momentum'' $p=a^2\dot x$ as effective velocity ($v_{\rm pec}=p/a$,)
and change the independent variable from time $t$ to expansion
parameter $a$, then the equations are 
$$\eqalign{
{d{\bf p}\over da} &= -{\nabla\phi\over\dot a}, \quad
{d{\bf x}\over da} =  {{\bf p}\over \dot a a^2}, \quad
\nabla^2\phi = 4\pi G\Omega_0a^2\delta_{\rm dm}, \cr
\dot a &= {H_0\over a^{1/2}}\sqrt{\Omega_0 
         +\Omega_{\rm curv,0}a
         +\Omega_{\Lambda,0}a^3},
	\quad \Omega_0+\Omega_{\rm curv,0}+ \Omega_{\Lambda,0} =1
}
\nxt{\Eqpart}$$
where $\Omega_0$, $\Omega_{\rm curv,0}$, and $\Omega_{\Lambda,0}$ are
the density of the matter, effective densities of the curvature and
cosmological constant in units of the critical density at $z=0$. The
curvature contribution is positive for negative curvature.  
\bigskip
\centerline{\bf 3. Methods}

There are many different numerical techniques to follow the evolution
of a system of many particles. For earlier reviews see Hockney \&
Eastwood \nr{\HockneyEast} and Sellwood \nr{\Sellwood}. The most
frequently used methods for cosmological applications fall in three
classes: Particle Mesh (PM) codes, Particle-Particle/Particle-Mesh
(P$^3$M) codes, and TREE codes. All methods have their advantages and
disadvantages.

{\bf PM code}. It uses a mesh to produce density and potential.  As
the result, its resolution is limited by the size of the mesh. Largest
simulations were done by the author: $800^3$ mesh with $3\times 256^3
=1.5\times 10^8$ particles. New parallel supercomputer SP2 at Cornell
will be used to run simulations with $1600^3=4.096\times 10^9$ mesh.
There are two advantages of the method: i) it is fast (the smallest
number of operations per particle per time step of all the other
methods), ii) it typically uses very large number of particles. The
later can be crucial for some applications. There are few
modifications of the code. There are few variants of PM code.
``plain-vanilla'' PM was described by Hockney \& Eastwood
\Ref{\HockneyEast}. It includes Cloud-In-Cell density assignment and
7-point discrete analog of the laplacian operator. Higher order
approximations improve the accuracy on large distances, but degrade
the resolution (e.g. \nr{\Gelb} ). In an effort to reduce the order of
approximation and to increase the resolution, Melott \nr{\Melott}
introduced staggered mesh. It gives a better resolution on cell-size
distances, but particles get self-forces (an isolated particle
experiences a force from itself), which might be not a welcome
feature. No serious testing of the method was done.

{\bf P$^3$M code} is described in detail in \Ref{\HockneyEast} and
\nr{\Efff}. It has two parts: PM part, which
takes care of large-scale forces, and PP part, which adds small-scale
particle-particle contribution. The simulations usually have
$64^3$--$100^3$ particles. The best known to me simulations were done
by Ma \& Bertschinger \nr{\Ma} for CHDM model: $128^3$ of cold
particles and ten times more hot particles. Because of strong
clustering at late stages of evolution, PP part becomes prohibitively
expensive once large objects start to form in large
numbers. Significant speed is achieved in modified version of the
code, which introduces subgrids (next levels of PM) in areas with high
density \nr{\Couchman}. With modification the code runs as fas as TREE
code even for heavily clustered configurations \Ref{\Couchman}.

{\bf TREE code} is the most flexible code in the sense of the choice
of boundary conditions [\nrr{\Appel}, \nrr{\BarnesHut},
\nrr{\Hernquist}. It is also more expensive than PM: it takes 10-50
times more operations.  Bouchet \& Hernquist \nr{\BouchetHernquis} and
Hernquist, Suto \& Bouchet \nr{\HSB} extended the code for the
periodical boundary conditions, which is important for simulating
large-scale fluctuations. An interesting blend of PM and TREE code as
done by Xu \nr{\Xu}. The code might be easier than usual to implement
on parallel supercomputers.

Multigrid methods were introduced long ago, but only recently they
started to show a potential to produce real results [\nrr{\Villumsen},
\nrr{\Anninos}, \nrr{\Splinter}].  It worth of paying attention if a
``multigrid'' code is really a fully adaptive multigrid code. Some of
the them are actually two-level mesh codes. Those codes provide
increased resolution (by factor 4--8) inside one predefined region.

What code is the best? Which one to choose? Should we through away
results of low resolution codes? (Tempting, but I would not do that).
There is no unique answer -- everything depends on the problem, which
we are addressing. For example, if we are interested in explanation of
the large-scale structure (filaments, voids, Zeldovich approximation,
and so on), PM code with 256$^3$ mesh is sufficient. It takes only one
night to make a simulation on a (good) workstation. There is a very
long list of problems like that.  But if you intent to look for
galaxies in the large-scale environment, much better resolution is
needed.  Sometimes a minimum dynamical range of $10^4$ is quoted: we
need to go to 10~kpc to see a galaxy, and we need to go up to at least
100~Mpc in order to have large voids and superclusters. What we should
also include in this ``wish-list'' is a required mass range. If we
would like to have something, which looks like a $10^{11}\Msun$
galaxy, it should consist of many particles. Say, 30-100
particles. This gives $10^9\Msun$ per particle and $410^3$ particles
for 100~Mpc box. All existing methods fall short of these
requirements: PM is lacking resolution (a factor of ten), others do
not have enough mass resolution (a factor of ten).

\bigskip
\centerline{\bf 4. Distribution of the matter and effects of resolution}

Figure 1 shows an example of structures observed in cosmological
models. The following model with cosmological constant was chosen:
$\Omega_0=0.30$, $h=0.70$, $\sigma_8=1.1$, $Q_{\rm rms-PS}=21.8\mu K$.
The simulation was done with the PM code with $256^3$ particles,
$800^3$ mesh, $\Delta a=0.003$, box size $50\Mpch$. Dark matter
particles in a $1\Mpch$ slice are shown. Different structures can be
identified on the plot.  Large empty (of galaxies) voids, long
filaments, few groups of galaxies, and many isolated galaxies. While
it is not clear from this plot whether we actually see filaments or
sheets, Kofman \etal \nr{\Kofman} recently argued that they actually
are filaments. It was a long-standing problem: the Zeldovich
approximation was predicting ``pancakes'' and what was ``numerically
observed'' looked more like filaments. The argument from the Zeldovich
approximation is still valid. If we smooth initial density field and
then randomly pick up a point at very early stage and trace its
trajectory, then the particle will first hit a sheet, not a filament,
because its middle eigenvalue of the deformation tensor is zero (on
average) and the smallest one is negative. (Positive sign of an
eigenvalue implies collapse along corresponding direction. Only one
positive eigenvalue means collapse along only one directions -- a
sheet forms). But Kofman \etal argue that this is not we are doing. We
observe particle at a given moment. In this case we are asking what is
the shape of an isodensity surface with density of few mean densities.
This condition results in positive middle eigenvalue while keeping the
expansion along the third axis -- this is a filament.

In order to discuss objects on smaller scales we need to understand
what happens with an isolated object, which size is close to the limit
of resolution. In all codes the resolution (either a cell size or a
softening parameter) is constant in {\it comoving} not proper
coordinates. This means that the resolution actually decreases with
time. (Look at the situation from a positive angle - the resolution
gets much better if we go to high redshifts).  For example, our
resolution is 50~kpc and we have a galaxy with radius 50~kpc
collapsing at $z=4, a=0.2$. The resolution at that redshift was 10~kpc
-- enough to reasonably resolve the object. Now, our resolution drops
very severely. If this happened quickly (on dynamical scale), the
galaxy would explode: its total energy would be positive.
Fortunately, this does not happen because the time-scale of growth of
the softening parameter is equal to the expansion rate of the
Universe, and the later is always longer than the dynamical time. Thus
the galaxy will adiabatically expand once the softening length gets
close to its radius. Figure 2 illustrates this behavior. A system of
300 particles with initial velocities slightly below virial velocities
was simulated using a PP code. Initially the particles were randomly
distributed in a sphere with 100~kpc radius. Parameters were scaled in
such a way that the total mass of the system is $10^{12}\Msun$,
initial moment corresponds to $a=0.2, z=4$. Collapse happens at
$a=0.25, z=3$. The radius of half mass is 50~kpc after collapse. The
system was run once with softening parameter $\epsilon =10$~kpc in
{\it proper} coordinates, and another time with $\epsilon =10$~kpc in
{\it comoving} coordinates.  Figure 2 shows snapshots of the
simulations. Comoving coordinates are chosen to display the
particles. Coordinate scale is in units of 100~kpc.  Bottom row shows
the evolution of the system simulated with constant proper resolution,
while the top row is for constant comoving resolution. The circles
show the size of resolution.  Radius of each circle is
$1.5\epsilon$. At this radius the error in force at 1 cell radius in
PM code is equal to the error in PP code. As time goes on the system
shrinks in comoving coordinates. But once its size is close to the
resolution, the shrinking stops.

This just confirms what is found in numerical simulations: isolated
unresolved ``galaxies'' are small, almost spherical balls with 2 grid
cells across their diameters.

In order get a better insight on the effects of the resolution and to
understand what should be expected if we significantly increase the
resolution, two simulations were run with exactly the same initial
conditions, but with twice different resolution. The same cosmological
model as before was chosen ($\Lambda$CDM with $\Omega_0=0.3, h=0.70$).
PM code with $128^3$ particles, mesh $450^3$ and $216^3$, box size
$20\Mpch$ was used. This gives the resolution of $44h^{-1}$kpc and
$92h^{-1}$kpc for the two runs and mass for a particle $m_1=3.1\times
10^8h^{-1}\Msun$. Figure 3 shows particles in a small 3x3~Mpc window
in the simulations at $z=3$. Only particles with estimated density
indicated in the figure are shown and only 1/4 of all particles is
displayed. Because the same isolated clump would have smaller radius
and, thus, higher density with twice better resolution, density limits
were slightly adjusted to take into account the difference in the
resolution. Figure 4 shows the same window at $z=0$. While plots for
higher resolution show more small clumps, all large clumps are found
in both plots. The differences are as expected: higher resolution
results in more compact and dense objects. The most significant result
lies in what was feared, but not found. The largest object in $z=0$
plot has radius of about 300~kpc. The usual question was: having a
low-resolution run with 100~kpc resolution, how do you know that with
better resolution it will not split into, say into two objects?  Our
results indicate that it does not happen: with more then twice better
resolution the object got a bit smaller, a bit rounder, got few tiny
satellites, which were barely seeing with the low resolution, but it
does not show any tendency to break int large peaces. Figures 5 and 6
show another examples of comparison of the two simulations. In this
case we deal with a group of galaxies. The result is just the same:
more small objects, the same large objects, no tendency for splitting
of large halos into smaller ones.

While visual analysis indicates that we might be quite ok with
relatively low resolution if only large halos (above $10^{10}\Msun$)
are considered, it is not that easy to make quantitative analysis.
Comparison of coordinates, velocities, and densities of individual
particles in both simulations would be a natural test, but it just
fails. The reason is the divergence of trajectories in dynamical
systems: small differences in initial coordinates result in large
differences after few dynamical times. The winding problem of the
spiral pattern is just one of the manifestation of this divergence.
As the result, deviations of coordinates $x_{44}-x_{92}$ and
velocities $V_{x,44}-V_{x,92}$ as functions of density (Figures 7 and
8) indicate mainly sizes of virialized object, which exist at
different redshift. Even in spite of the divergence of the
trajectories, extremely large fraction of particles -- 99\% --
indicate the difference in coordinates less than 2 cell sizes and
difference in velocities less than 100~km/s.
\bigskip
\centerline{\bf 5. Halo identification and overmerging problem}

There are different methods of identifying collapsed 
objects (halos) in numerical simulations. 

{\bf Friends-Of-Friends (FOF)} algorithm was used a lot and still has
its adepts. If we imagine that each particle is surrounded by a sphere
of radius $b d/2$, then every connected group of particles is
identified as a halo. Here $d$ is the mean distance between particles,
and $b$ is called {\it linking parameter}, which typically is 0.2.
Dependance of groups on $b$ is extremely strong. The method stems from
an old idea to use percolation theory to discriminate between
cosmological models.  Because of that, FOF is also called percolation
method, which is wrong because the percolation is about groups
spanning the whole box, not collapsed and compact objects. FOF was
criticized for failing to find separate groups in cases when those
groups were obviously present
\Ref{\Gelb}. The problem originates from the tendency of FOF to 
``percolate'' through bridges connecting interacting galaxies or
galaxies in high density backgrounds. 

{\bf DENMAX} tried to overcome the problems of FOF by dealing with
density maxima [\ref{\Gelb},\nrr{\GelbBert}]. It finds maxima of
density and then tries to identify particles, which belong to each
maximum (halo). The procedure is quite complicated. First, density
field is constructed. Second, the density (with negative sign) is
treated as potential in which particles start to move as in a viscous
fluid.  Eventially, particles sink at bottoms of the potential (which
are also maxima density). Third, only particles with negative energy
(relative to their group) are retained. Just as in the case of FOF, we
can easily imagine situations when (this time) DENMAX should fail. For
example, two colliding galaxies in a cluster of galaxies. Because of
large relative velocity they should just pass each other. In the
moment of collision DENMAX ceases to ``see'' both galaxies because all
particle have positive energies. That is probably quite unlikely
situation. The method is definitely one of the best at present. The
only problem is that it seems to be too complicated for present state
of simulations.
 
{\bf ``Overdensity 200''}. There is no name for the method, but it is
often used. Find density maximum, place a sphere and find radius,
within which the sphere has the mean overdensity 200 (or 177 if you
really want to follow the top-hat model of nonlinear collapse). In
Figure 5 all particles in the central $\sim$1~Mpc area will be just
one halo. The mass function of dark halos, constructed in this way,
has a very long tail extending into masses, which are far too large
for individual galaxies. The existence of this tail is often called
``overmerging problem''. Figure 9 gives comparison of mass functions
of dark halos $n(>M)$ in previous two simulations with different
resolutions. The full curve shows the mass function for high
resolution run. Note that high mass tail of the mass functions does
not depend on resolution, which indicates convergence of the results.
The mass function flattens in both simulations at different level
(more small-mass objects with high resolution), but at the same mass,
which corresponds to 20-30 particles per object.

Actually, there are two ``overmerging'' problems. One arises because
of naive application of the top-hat model. Consider our Galaxy and
Andromeda Nebula as a testbed. If we assume that both galaxies have
flat rotation curves at least up to the radius of 100~kpc (which might
be true), then overdensity at 100~kpc is about 4000 for our Galaxy and
about twice that for Andromeda Nebula. The overdensity at half way
between galaxies is about 200--300. Thus, if we had fantastic
simulation with our Galaxy and M31 as they are in the Universe, our
``overdensity 200'' method would lump them together. There are
different ways of fixing the problem. DENMAX probably would find both
galaxies.

Unfortunately, there is real ``overmerger'' problem, which cannot be
avoided that easily. For a long time it was assumed that appearance of
extremely large halos (``overmergers'') is due to insufficient
resolution. Numerical simulations indicated that with smaller and
smaller mass of each particle, and with better and better resolution
we see more and more substructure in overmergers. But it seems that
the tendency have stopped. Now extra resolution does not split large
overmergers. The problem is real and it is not a numerical artifact,
which can be cured by much better resolution. Figures 3-6 illustrate
this result. Recently, Moore \etal \nr{\Moore} and van Kampen
\nr{\Kampen} gave a very plausible explanation: tidal forces.  If a
small halo (``galaxy'') falls into a larger one (``group''), it will
be tidally disrupted at a distance where its central density is equal
to the mean density of the large halo at that distance (for ``group''
with $\rho\propto r^{-2}$). Because the core of the large halo likely
was formed earlier than the small halo, it has higher density. The
only chance for the small halo to survive is to stay in peripheral
parts of the group. Numerical simulations show that this is what
happens.

\vfill\eject
\centerline{\bf References}
\parindent = 0pt
\parskip=0pt plus4pt minus2pt

\ps \Reff{\Aarseth} Aarseth S.J., in {\it Multiple Time Scales} edited by
	J. W. Brackbill and B. J. Cohen (New York, Academic) 1985, p. 377.

\ps \Reff{\Peeblesa} Peebles P.J.E., \aj {\bf 75} (1970) 13.

\ps \Reff{\GunnGott} Gunn J.E. and Gott J.R., \apj {\bf 176} (1972) 1.

\ps \Reff{\Whitea} White S.D.M., \mnras {\bf 177} (1976) 717.

\ps \Reff{\AarsethGT} Aarseth S.J., Gott J.R., and turner E.L., {\bf 228}
	(1979) 664.          

\ps \Reff{\Peeblesb} Peebles P.J.E., {\it The large-scale structure
	of the universe}, Princeton University Press  (1980). 

\ps \Reff{\Zeldovich} Zeldovich Ya.B., \aa {\bf 5} (1970) 84.

\ps \Reff{\EfEastwood} Efstathiou G. and Eastwood J.W., \mnras {\bf 194}
	(1981) 503.

\ps \Reff{\KlypinShandarin} Klypin A. and Shandarin S.F., \mnras{\bf 204}
	(1983) 891. 

\ps \Reff{\Doroshkevich} Doroshkevich A.G., Kotok E.V., Novikov I.D.,
	Polyudov A.N. and Sigov Yu.S., \mnras {\bf 192 } (1980) 321.

\ps \Reff{\PS} Press W.H. and Schechter P., \apj {\bf 187} (1974) 425.

\ps \Reff{\ColesMS} Coles P., Melott A.L., and Shandarin S.F., \mnras {\bf
	260} (1993) 765 .

\ps \Reff{\Sath} Sathyaprakash B.S., Sahni V., Munshi D., Pogosyan D., and
	Melott A.L., \mnras  (1995) submitted.

\ps \Reff{\BCM} Borgani S., Coles P., Moscardini L., \mnras {\bf 271}(1994)223.

\ps \Reff{\EfRees} Efstathiou G. and ReesM., \mnras {\bf 230} (1988) 5P.

\ps \Reff{\KHPR} Klypin A., Holtzman J., Primack J., and Regos E., \apj
	{\bf 416} (1993) 1. 

\ps \Reff{\HockneyEast} Hockney R.W. and Eastwood J.W., {\it Numerical
        simulations using particles} (New York: McGraw-Hill) 1981. 

\ps \Reff{\Sellwood} Sellwood J.A., {\it Ann. Rev. Astron. Astrophys.}
	{\bf 25} (1987) 151. 

\ps \Reff{\Gelb} Gelb J., {\it Ph.D. Thesis,} MIT {\bf } (1992).

\ps \Reff{\Melott} Melott A.L., {\it Phys. Rev. Letters} {\bf 56} (1986) 1992.

\ps \Reff{\Efff} Efstathiou G., Davis M., Frenk C.S., and  White S.D.M.,
	\apjs {\bf 57} (1985) 241.

\ps \Reff{\Ma} Ma C. and Bertschinger E., \apj  {\bf 434 } (1994) L5.

\ps \Reff{\Couchman}  Couchman H.M.P., \apj  {\bf 368 } (1991) 23.

\ps \Reff{\Appel} Appel A., {\it SIAM J. Sci. Stat. Comput.}, {\bf 6 }
	(1985) 85.

\ps \Reff{\BarnesHut} Barnes J. and Hut P., {\it Nature} {\bf 324} (1986) 446.

\ps \Reff{\Hernquist} Hernquist L., \apjs  {\bf 64 } (1987) 715.

\ps \Reff{\BouchetHernquis} Bouchet F.R. and Hernquist L., \apjs {\bf 68 }
	(1988) 521. 

\ps \Reff{\HSB} Hernquist L., Bouchet F.R., and Suto Y., \apjs {\bf 75 }
	(1991) 231. 

\ps \Reff{\Xu}  Xu G., \apj  {\bf 98 } (1995) 355.

\ps \Reff{\Villumsen} Villumsen J.V., \apjs {\bf 71} (1988) 407.

\ps \Reff{\Anninos} Aninnos P., Norman M., Clarke D.A., \apj {\bf 436}
	(1994) 11.

\ps \Reff{\Splinter} Splinter R.J., \mnras in press.

\ps \Reff{\Kofman} Kofman L., Pogosian D., Bond J.R., and Klypin A., \apj
	(1995) submitted. 

\ps \Reff{\GelbBert} Bertschinger E. and Gelb J., {\it Comp. Phys.} {\bf 5}
	(1991) 164.

\ps \Reff{\Moore} Moore B., Katz N., and Lake, G., \apj (1995) submitted.

\ps \Reff{\Kampen} van Kampen E., \mnras {\bf 273} (1995) 295. 

%\ps \Reff{}
\bigskip
\centerline{\bf Figure Captions}

{\bf Figure 1} An example of structures observed in cosmological
models with cosmological constant:
$\Omega_0=0.30$, $h=0.70$, $\sigma_8=1.1$, $Q_{\rm rms-PS}=21.8\mu K$.
The simulation was done with the PM code with $256^3$ particles,
$800^3$ mesh. Dark matter
particles in a $1\Mpch$ slice are shown.

{\bf Figure 2} Evolution of a system of 300
particles in comoving coordinates. Initially the particles were randomly
distributed in a sphere with 100~kpc radius. Collapse happens at
$a=0.25, z=3$.  The
system was run once with softening parameter 
$\epsilon =10$~kpc in {\it proper} coordinates (bottom row), and another
time with  $\epsilon =10$~kpc in {\it comoving} coordinates (top row).
Coordinate scale is in units of 100~kpc. The circles show the size of
resolution. Radius of each circle is $1.5\epsilon$. 

{\bf Figure 3} Particles in a small 3x3~Mpc window
 at $z=3$ in $\Lambda$CDM model with  $\Omega_0=0.3, h=0.70$. Only
particles with estimated density 
indicated in the figure are shown and only 1/4 of all particles is
displayed.

{\bf Figure 4} The same window at $z=0$.

{\bf Figure 5} An example of a group of galaxies simulated with with
different resolutions.

{\bf Figure 6} The sane as in Figure 5, but with different density
threshold. 

{\bf Figure 7} Deviations of coordinates and velocities in high and
low resolution simulations at $z=3$.

{\bf Figure 8} The same as in Figure 7, but at $z=0$.

{\bf Figure 9} Comparison of mass
functions of dark halos $n(>M)$ in two simulations with
different reolutions ($44h^{-1}$~kpc and $92h^{-1}$~kpc). The full
curve shows the mass function for high 
resolution run. Note that high mass tail of the mass functions does
not depend on resolution, which indicates convergence of the results.

\bye